\documentclass[aps,11pt,nofootinbib, tightenlines]{revtex4-2}
\pdfoutput=1
\usepackage[utf8]{inputenc}
\usepackage[T1]{fontenc} 
\usepackage{lmodern} 
\usepackage[margin=1in]{geometry}
\usepackage[all]{nowidow}

\usepackage{graphicx}
\usepackage{float}

\usepackage{amsmath}
\usepackage{amssymb}
\usepackage{bm}
\usepackage{multirow}

\newcommand{\PBH}{\mathrm{PBH}}

\newcommand{\diffd}{\mathrm{d}} 
\newcommand{\ddv}[2]{\frac{\diffd #1}{\diffd #2}} 
\newcommand{\pdv}[2]{\frac{\partial #1}{\partial #2}} 
\newcommand{\vV}{\bm{V}}
\newcommand{\vSigma}{\bm{\Sigma}}
\newcommand{\vx}{\bm{x}}

\newcommand{\G}{\mathrm{G}}

\newcommand{\cdN}{classical USR $\delta N$}

\newcommand{\ie}{\textsl{i.e.~}}

\newcommand{\eg}{\textsl{e.g.~}}

\newcommand{\etc}{\textsl{etc.~}}

\newcommand{\fnl}{f_{\mathrm{NL}}}
\newcommand{\Eq}[1]{eq.~\eqref{#1}}

\makeatletter
\DeclareRobustCommand*\uell{{\mathpalette\@uell\relax}}
\newcommand*\@uell[2]{
	\setbox0=\hbox{$#1\ell$}
	\setbox1=\hbox{\rotatebox{10}{$#1\ell$}}
	\dimen0=\wd0 \advance\dimen0 by -\wd1 \divide\dimen0 by 2
	\mathord{\lower 0.1ex \hbox{\kern\dimen0\unhbox1\kern\dimen0}}
}
\makeatother

\usepackage[linktocpage=true,colorlinks=true,linkcolor=blue,citecolor=blue,urlcolor=blue]{hyperref}
\hypersetup{pdftitle={Non-perturbative non-Gaussianity and primordial black holes},pdfauthor={5 authors}}

\begin{document}

\author{Andrew D.~Gow$^{1}$}
\email{andrew.gow@port.ac.uk}

\author{Hooshyar Assadullahi$^{1,2}$}
\email{hooshyar.assadullahi@port.ac.uk}

\author{Joseph H.~P.~Jackson$^{1}$}
\email{joseph.jackson@port.ac.uk}

\author{Kazuya Koyama$^{1}$}
\email{kazuya.koyama@port.ac.uk}

\author{Vincent Vennin$^{3,4,1}$}
\email{vincent.vennin@ens.fr}

\author{David Wands$^{1}$}
\email{david.wands@port.ac.uk}

\affiliation{\\1) Institute of Cosmology \& Gravitation, University of Portsmouth, \mbox{Dennis Sciama Building, Burnaby Road}, Portsmouth, PO1 3FX, United Kingdom\\}

\affiliation{\\2) School of Mathematics and Physics, University of Portsmouth, Lion Gate Building, Lion Terrace, Portsmouth, PO1 3HF, United Kingdom\\}

\affiliation{\\3) Laboratoire de Physique de l'Ecole Normale Sup\'erieure, ENS, CNRS, Universit\'e PSL, Sorbonne Universit\'e, Universit\'e Paris Cit\'e, 75005 Paris, France\\}

\affiliation{\\4) Laboratoire Astroparticule et Cosmologie, CNRS, Universit\'e Paris Cit\'e, 75013 Paris, France\\}

\date{31/05/2023}

\title{Non-perturbative non-Gaussianity and primordial black holes}

\begin{abstract}
We present a non-perturbative method for calculating the abundance of primordial black holes given an arbitrary one-point probability distribution function for the primordial curvature perturbation, $P(\zeta)$. A non-perturbative method is essential when considering non-Gaussianities that cannot be treated using a conventional perturbative expansion. To determine the full statistics of the density field, we relate $\zeta$ to a Gaussian field by equating the cumulative distribution functions. We consider two examples: a specific local-type non-Gaussian distribution arising from ultra slow roll models, and a general piecewise model for $P(\zeta)$ with an exponential tail. We demonstrate that the enhancement of primordial black hole formation is due to the intermediate regime, rather than the far tail. We also show that non-Gaussianity can have a significant impact on the shape of the primordial black hole mass distribution.
\end{abstract}

\maketitle

\section{Introduction}
One of the unanswered questions in $\Lambda$CDM cosmology is the nature of dark matter. One suggestion that has attracted renewed interest recently is that of primordial black holes (PBHs), which could have formed from the collapse of very overdense regions in the early universe \cite{Zel'dovich:1967_Cores,Hawking:1971_Gravitationally,Carr:1974_Black}. If they survive to the present day, they would constitute some or all of the dark matter content of the universe. This is characterised through the parameter $f_\PBH$, the fraction of dark matter in the form of PBHs, \ie $f_\PBH = 1$ indicates that PBHs make up the entirety of the dark matter. For a recent review of constraints on $f_\PBH$, see Ref.~\cite{Carr:2020_Constraints}. 
Even if they are not the main constituent of dark matter, the merger of PBH binaries will also generate gravitational waves, and may contribute to the LIGO--Virgo--KAGRA detections~\cite{Bird:2016_GW,Sasaki:GW,Clesse:2017_Clustering,Gow:2020_ACAJ,Franciolini:2021_GW}.

The simplest formation mechanism for PBHs is from overdensities seeded by inflation, a period of accelerated expansion in the very early universe that also explains the seeding of large scale structure. The perturbations from inflation are typically characterised by the primordial power spectrum of the curvature perturbation, $\mathcal{P}_\zeta(k)$, which is observed on cosmic microwave background (CMB) scales to be $\sim2.1\times10^{-9}$ \cite{Planck:2018_Parameters}. 
It is commonly assumed that to form PBHs requires increasing $\mathcal{P}_\zeta(k)$ by several orders of magnitude on small scales~\cite{Gow:2021_ACPS}. However, this assumes that the probability distribution function (PDF) of the curvature perturbation is Gaussian. PBHs are formed in the tail of the PDF, and so non-Gaussianity that enhances the occurrence of rare events could have a significant impact on PBH formation, potentially reducing the power required to form them.

Previous studies of PBH abundance in the presence of primordial non-Gaussianity have usually assumed a local, perturbative expansion in terms of a first-order Gaussian curvature perturbation, $\zeta_{\G}$ with higher-order parameters $\fnl$, \etc\cite{Bullock:1996_Non-Gaussian,Young:2015_Influence,Yoo:2019_Abundance,Palma:2020_Non-Gaussian,Taoso:2021_Non-gaussianities,Young:2022_non-G}. However, a non-perturbative calculation can be made using the $\delta N$-method~\cite{Lyth:2005_Inflationary}. It has been argued that by fully incorporating quantum diffusion, in the so-called stochastic $\delta N$ formalism~\cite{Fujita:2013_Algorithm, Vennin:2015_Correlation}, we expect the far tail of the probability distribution for primordial curvature perturbations to generically display a non-Gaussian, exponential tail~\cite{Pattison:2017_Quantum,Ezquiaga:2019_Exponential,Vennin:2020_Thesis,Figueroa:2020_Non-Gaussian,Ando:2020_Power,Pattison:2021_USR,Rigopoulos:2021_Inflation, Tada:2021_Statistics,Jackson:2022_Numerical,Animali:2022_Primordial} (such heavy tails were also found in Refs.~\cite{Panagopoulos:2019_Primordial, Achucarro:2021_Hand-made, Cai:2022_Highly} using different methods). Therefore it is important to develop and apply techniques which enable us to explore more general forms for the PDF, and for the tail of the distribution in particular, which encompass a wider range of non-Gaussianity. In this paper, we present for the first time a fully non-perturbative treatment of primordial non-Gaussianity that makes no assumptions about the model other than that $\zeta$ is related to a Gaussian field $\zeta_{\G}$ by what we will call a ``generalised local'' transformation.

\section{Method}
\label{sec:Method}

Primordial black holes form from rare, large fluctuations in the primordial density field. The power spectrum is no longer sufficient to determine the abundance of large fluctuations if we wish to go beyond the simplest Gaussian distribution. Instead we will start from an arbitrary PDF for the primordial curvature perturbation $P[\zeta(\vx)]$, \textsl{e.g.}, obtained from an inflationary model.

The criteria for formation of PBHs can be given in terms of the compaction function, $C(r)$. This is a volume-averaged version of the density contrast, equivalent to smoothing with a real-space top-hat window function~\cite{Shibata:1999_Compaction,Harada:2015_Compaction}. The relationship between $\zeta$ and $C$ is non-linear, but the compaction during radiation domination can be written in terms of its linear part $C_{\uell}$ as
\begin{align}
C &= C_\uell - \frac{3}{8}C_\uell^2, \label{eq:C}
\end{align}
with $C_\uell$ related to $\zeta$ by
\begin{align}
C_\uell &= -\frac{4}{3}r\zeta', \label{eq:C_l}
\end{align}
where a prime $'$ denotes a derivative with respect to $r$. This simple relation makes it useful to calculate the PBH properties in terms of $P(C_\uell)$ instead of $P(C)$. Eq.~\eqref{eq:C} has a maximum of $C = 2/3$ for $C_\uell = 4/3$, which corresponds to the boundary between type I and type II perturbations. It is unclear what the properties of a PBH resulting from a type II perturbation would be~\cite{Kopp:2010_Separate}, so we restrict to $C_\uell \leqslant 4/3$ in this work.

For a Gaussian field, the high peaks in the density field that form PBHs can be approximated as spherical. We assume that this also holds in our non-Gaussian case, and we will show that this is a self-consistent assumption in our subsequent calculations. For the linear compaction in \Eq{eq:C_l} to be spherical, the curvature field $\zeta$ in the same region must also be spherical. In general, we can expand the field in spherical harmonics,
\begin{align}
\zeta(\vx) &= \zeta_{00}(r) + \sum_{\ell>0,m\in[-\ell,\ell]} Y_{\ell m}(\theta,\phi)\zeta_{\ell m}(r).
\end{align}
From now on we will consider only the spherically symmetric monopole term in this expansion and use the shorthand $\zeta \equiv \zeta(r) \equiv \zeta_{00}(r)$.

The reliance on $\zeta'$ rather than $\zeta$ itself means that $P(\zeta)$ is not sufficient to calculate the PBH mass distribution and abundance, and instead the joint probability $P[\zeta(r_1),\cdots,\zeta(r_n)]$ is required~\cite{DeLuca:2022_Note}. Therefore, in our analysis, we will assume that $\zeta(r)$ can be related to a Gaussian field $\zeta_{\G}(r)$ whose statistical properties are well-understood. We introduce a ``generalised local'' transformation,
\begin{align}
\zeta(r) &= \zeta[\zeta_{\G}(r),r],
\end{align}
where we have allowed for an explicit $r$-dependence independent of the Gaussian field $\zeta_{\G}$. This extra dependence is required to consider arbitrary $P(\zeta)$ since the Gaussian PDF will always carry an $r$ dependence through the variance $\Sigma_{YY}$ as well as $\zeta_G$, as can be seen in eqs.~\eqref{eq:zeta-transform-general}~and~\eqref{eq:zeta-transform}. This transformation is a generalisation of a local transformation, in the sense that it includes the local case, where $\zeta$ is a local function of a Gaussian field, $\zeta = \zeta[\zeta_{\G}(r)]$. For a monotonic transformation, high peaks in $\zeta$ necessarily correspond to high peaks in $\zeta_{\G}$, and hence this relation is consistent with our spherical assumption.

We want to connect an arbitrary non-Gaussian $\zeta$ with PDF $P(\zeta)$ to a Gaussian $\zeta_{\G}$ with PDF
\begin{align}
P_{\G}(\zeta_{\G},\Sigma_{YY}) &= \frac{1}{\sqrt{2\pi\Sigma_{YY}}}\exp\left(-\frac{\zeta_{\G}^2}{2\Sigma_{YY}}\right),
\end{align}
where we write the variance as $\Sigma_{YY}$ for reasons that shall become apparent. To do this, we utilise the fact that a variable defined by $F_X(X)$ is always uniformly distributed, where $F_X$ is the cumulative distribution function (CDF) for the variable $X$. This means we can equate the CDFs for $\zeta$ and $\zeta_{\G}$,
\begin{align}
F\left[\zeta(r), r\right] &= F_{\G}\left[\zeta_{\G}(r), \Sigma_{YY}(r)\right], \label{eq:zeta-transform-general}
\end{align}
which can be applied for any $P(\zeta)$, whether analytical or numerical.

The simplest generalised local transformation has all the $r$-dependence coming from $\zeta_{\G}(r)$ and the variance $\Sigma_{YY}(r)$, for which the non-Gaussian field can be written as
\begin{align}
\zeta(\zeta_{\G},\Sigma_{YY}) &= F^{-1}[F_{\G}(\zeta_{\G},\Sigma_{YY})], \label{eq:zeta-transform}
\end{align}
where $F^{-1}$ is the inverse CDF for $\zeta$. In this case, $\zeta'$ in \Eq{eq:C_l} includes two terms, one from the dependence of $\zeta_{\G}$ on $r$, and one from the dependence of $\Sigma_{YY}$ on $r$. We can therefore write the linear compaction~\eqref{eq:C_l} as
\begin{align}
C_\uell &= -\frac{4}{3}r\left[\mathcal{J}_1(\zeta_{\G})\zeta_{\G}' + \mathcal{J}_2(\zeta_{\G})\Sigma_{YY}'\right], \label{eq:C_l-zz'}
\end{align}
with Jacobian factors given by
\begin{align}
\mathcal{J}_1(\zeta_{\G}) &= \pdv{\zeta}{\zeta_{\G}}, \quad
\mathcal{J}_2(\zeta_{\G}) = \pdv{\zeta}{\Sigma_{YY}}.
\end{align}
Note that by construction, from \Eq{eq:zeta-transform}, we have $\mathcal{J}_1 = P_{\G}(\zeta_{\G})/P(\zeta)$. In the case of a local transformation where $\zeta$ depends only on $\zeta_{\G}$, $\mathcal{J}_2$ will be zero.

We can see from \Eq{eq:C_l-zz'} that the linear compaction depends on both $\zeta_{\G}$ and $\zeta_{\G}'$, and so $P(C_\uell)$ will be written in general as an integration over a 2D PDF,
\begin{align}
P(C_\uell) &= \int \diffd X \int \diffd Y\ P(X,Y) \delta_{\mathrm{D}}\left[C_\uell - C_\uell(X,Y)\right], \label{eq:P(C_l)-XY}
\end{align}
where $C_\uell(X,Y)$ is given by \Eq{eq:C_l-zz'}, $\delta_{\mathrm{D}}$ is a Dirac delta function, and $X$ and $Y$ are given by
\begin{align}
X &= r\zeta_{\G}', \quad Y = \zeta_{\G}.
\end{align}
The 2D Gaussian distribution, $P(X,Y)$, is given by
\begin{align}
P(X,Y) &= \frac{1}{2\pi\sqrt{\det(\vSigma)}}\exp\left(-\frac{\vV^T \vSigma^{-1} \vV}{2}\right), \label{eq:P(XY)}
\end{align}
with
\begin{align}
\vV^T &= (X,Y), \quad
\vSigma = \begin{pmatrix}
\Sigma_{XX} & \Sigma_{XY} \\
\Sigma_{XY} & \Sigma_{YY}
\end{pmatrix}.
\end{align}
The correlators can then be written in terms of the power spectrum of the Gaussian field, $\mathcal{P}_{\zeta_{\G}}(k)$,
\begin{align}
\Sigma_{XX} &= \int \diffd (\ln k)\ (kr)^2 \left[\ddv{j_0}{z}(kr)\right]^2 \mathcal{P}_{\zeta_{\G}}(k), \\
\Sigma_{XY} &= \int \diffd (\ln k)\ (kr)\  j_0(kr)\ddv{j_0}{z}(kr) \mathcal{P}_{\zeta_{\G}}(k), \\
\Sigma_{YY} &= \int \diffd (\ln k)\ j_0^2(kr) \mathcal{P}_{\zeta_{\G}}(k),
\end{align}
where $j_0 = (\sin z)/z$ is the spherical Bessel function. Here we differ from Ref.~\cite{Biagetti:2021_Formation} by retaining the $\Sigma_{XY}$ term, which is not zero for isotropy as claimed. Indeed, for a monochromatic Gaussian power spectrum, studied in Ref.~\cite{Kitajima:2021_Primordial}, the fields are fully correlated, with $\Sigma_{XY}^2 = \Sigma_{XX}\Sigma_{YY}$, and the joint PDF in \Eq{eq:P(XY)} collapses to a 1D case. In this work, to retain generality, we keep the cross-correlation term but assume a lognormal shape for the Gaussian power spectrum,
\begin{align}
\mathcal{P}_{\zeta_{\G}}(k) &= \frac{A}{\sqrt{2\pi}\Delta}\exp\left[-\frac{\ln^2(k/k_*)}{2\Delta^2}\right],
\label{eq:PzGk}
\end{align}
with amplitude $A$, width $\Delta$, and peak scale $k_*$. This is normalised such that it approaches the monochromatic case of a Dirac delta in ln $k$ in the limit $\Delta \to 0$.

We can then rewrite the linear compaction in terms of $X$ and $Y$ as
\begin{align}
C_\uell &= -\frac{4}{3}\left[\mathcal{J}_1(Y)X + \mathcal{J}_2(Y)r\Sigma_{YY}'\right] \\
&= -\frac{4}{3}\left[\mathcal{J}_1(Y)X + 2\mathcal{J}_2(Y)\Sigma_{XY}\right]. \label{eq:C_l-XY}
\end{align}
Finally, carrying out the marginalisation in \Eq{eq:P(C_l)-XY} with $C_\uell(X,Y)$ given by \Eq{eq:C_l-XY},
\begin{align}
P(C_\uell) &= \int \diffd\zeta_{\G}\ \frac{3}{4|\mathcal{J}_1(\zeta_{\G})|} P\left\lbrace -\frac{1}{\mathcal{J}_1(\zeta_{\G})}\left[\frac{3}{4}C_\uell + 2\Sigma_{XY}\mathcal{J}_2(\zeta_{\G})\right], \zeta_{\G}\right\rbrace. \label{eq:P(C_l)-main}
\end{align}

The PBH mass $m$ is related to the linear compaction $C_\uell$ by the critical collapse relation~\cite{Choptuik:1992_Critical-collapse,Evans:1994_Critical-collapse,Niemeyer:1998_Critical-collapse}
\begin{align}
m &= KM_H\left(C_\uell - \frac{3}{8}C_\uell^2 - C_{\mathrm{c}}\right)^\gamma. \label{eq:Critical-collapse}
\end{align}
where $K$, $C_c$, and $\gamma$ are parameters that control the collapse, which in general are dependent on the shape of the perturbations~\cite{Musco:2018_CC,Young:2019_CC,Germani:2018_CC,Germani:2019_Nonlinear,Escriva:2020_CC}. The mass distribution is then given by~\cite{Kitajima:2021_Primordial}
\begin{align}
f(m) &= Z\frac{K\left(C_\uell - \frac{3}{8}C_\uell^2 - C_{\mathrm{c}}\right)^{\gamma + 1}}{\gamma\left(1 - \frac{3}{4}C_\uell\right)}P(C_\uell), \label{eq:f(m)}
\end{align}
where $Z$ depends on the peak scale in the power spectrum and the background cosmology. The diverging term in the denominator arises from the Jacobian factor when changing variable from $C_\uell$ to $m$, and is ultimately inherited from the non-linear form of $C$ in \Eq{eq:C}. The boundary between type I and type II perturbations at $C_\uell = 4/3$ corresponds to the maximum value of $C = 2/3$, and hence a zero derivative. The PBH abundance is determined from the mass distribution $f(m)$ through
\begin{align}
f_\PBH &= \int_{0}^{m_\text{max}} \diffd (\ln m)\ f(m), \label{eq:f_PBH}
\end{align}
where $m_\textsl{max}$ corresponds to $C_\uell = 4/3$.

\section{Results}

We now apply the method described in sec.~\ref{sec:Method} to two examples of non-Gaussian distributions, $P(\zeta)$. We choose the power spectrum width $\Delta = 0.3$ in \Eq{eq:PzGk}, and take the same peak scale as~\cite{Kitajima:2021_Primordial}, $k_* = 1.56\times10^{13}$ Mpc$^{-1}$. The scale $r$ is chosen such that $k_*r = 2.74$, which maximises the compaction function for a monochromatic power spectrum~\cite{Musco:2020_274}. Additionally we choose the same critical collapse parameters as Ref.~\cite{Kitajima:2021_Primordial}, $K = 1$, $C_{\mathrm{c}} = 0.587$, $\gamma = 0.36$\footnote{We note that the maximum scale and critical collapse parameters are valid for a monochromatic power spectrum, and will differ for our case of a lognormal with finite width~\cite{Musco:2018_CC,Young:2019_CC,Germani:2018_CC,Germani:2019_Nonlinear,Escriva:2020_CC}. However, the results of the non-Gaussian calculation presented here will not be strongly sensitive to these modifications.}, and write the PBH mass in terms of the horizon mass $M_H$ evaluated at the peak scale in the power spectrum, $M_{k_*}$. For these choices, the factor in the mass distribution (\ref{eq:f(m)}) is $Z = 6.88\times10^{16}$.

The first non-Gaussian distribution $P(\zeta)$ is an example of a local transformation arising from the classical $\delta N$ approach for an ultra slow roll (USR) phase in the early universe. This results in the transformation~\cite{Cai:2018_Revisiting, Atal:2019_Primordial,Pi:2021_Primordial, Biagetti:2021_Formation,Kitajima:2021_Primordial}
\begin{align}
\zeta = -\frac13 \ln \left(1-3\zeta_{\G}\right), \label{eq:USRzeta}
\end{align}
and in the non-Gaussian PDF
\begin{align}
P(\zeta) = \frac{1}{\sqrt{2\pi\Sigma_{YY}}} \exp \left[-\frac{(1-e^{-3\zeta})^2}{18\Sigma_{YY}}-3\zeta\right], \label{eq:USRPzeta}
\end{align}
where $\Sigma_{YY}$ is the variance of the Gaussian field $\zeta_{\G}$. From now on we shall refer to this as the \cdN\ approach. In this case, and assuming a monochromatic power spectrum, our general method reduces to the ratio distribution approach in Ref.~\cite{Kitajima:2021_Primordial}.

As an example of a minimal generalised local transformation governed by \Eq{eq:zeta-transform}, we also introduce a test model for $P(\zeta)$ consisting of a piecewise matching between a Gaussian distribution and a simple exponential tail,
\begin{align}
P(\zeta) &= N\begin{cases}
P_{\G}(\zeta) & \zeta \leqslant \zeta_{\mathrm{t}} \\
P_{\G}(\zeta_{\mathrm{t}})\exp[-\alpha(\zeta-\zeta_{\mathrm{t}})] & \zeta > \zeta_{\mathrm{t}}
\end{cases}. \label{eq:P(zeta)-piecewise}
\end{align}
The width of the Gaussian part is chosen to be identical to the distribution for $\zeta_{\G}$, and the factor $N$ is included to give the correct overall normalisation.

\begin{figure}[H]
\centering
\includegraphics[width=0.5\textwidth]{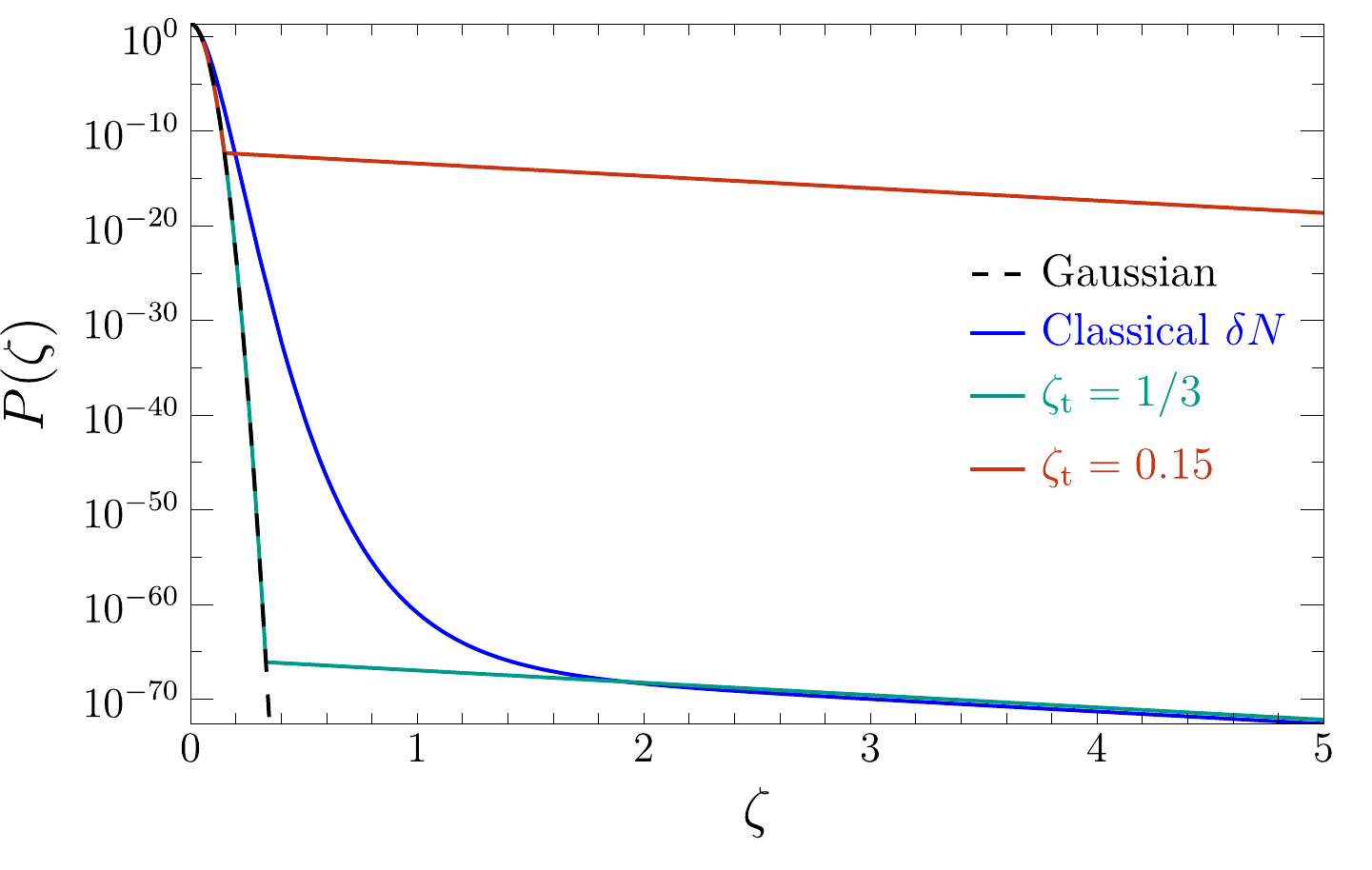}\includegraphics[width=0.5\textwidth]{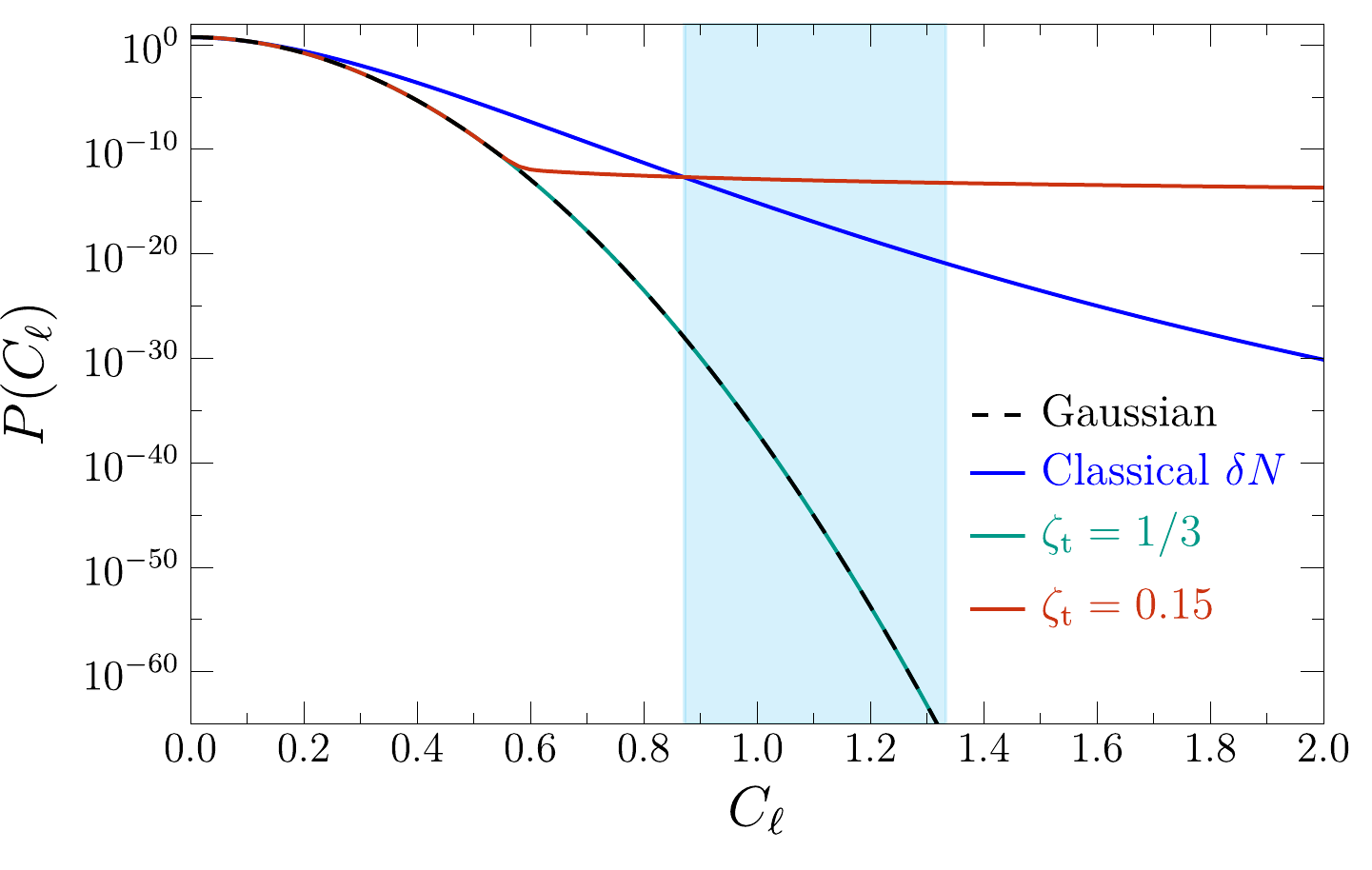}
\includegraphics[width=0.5\textwidth]{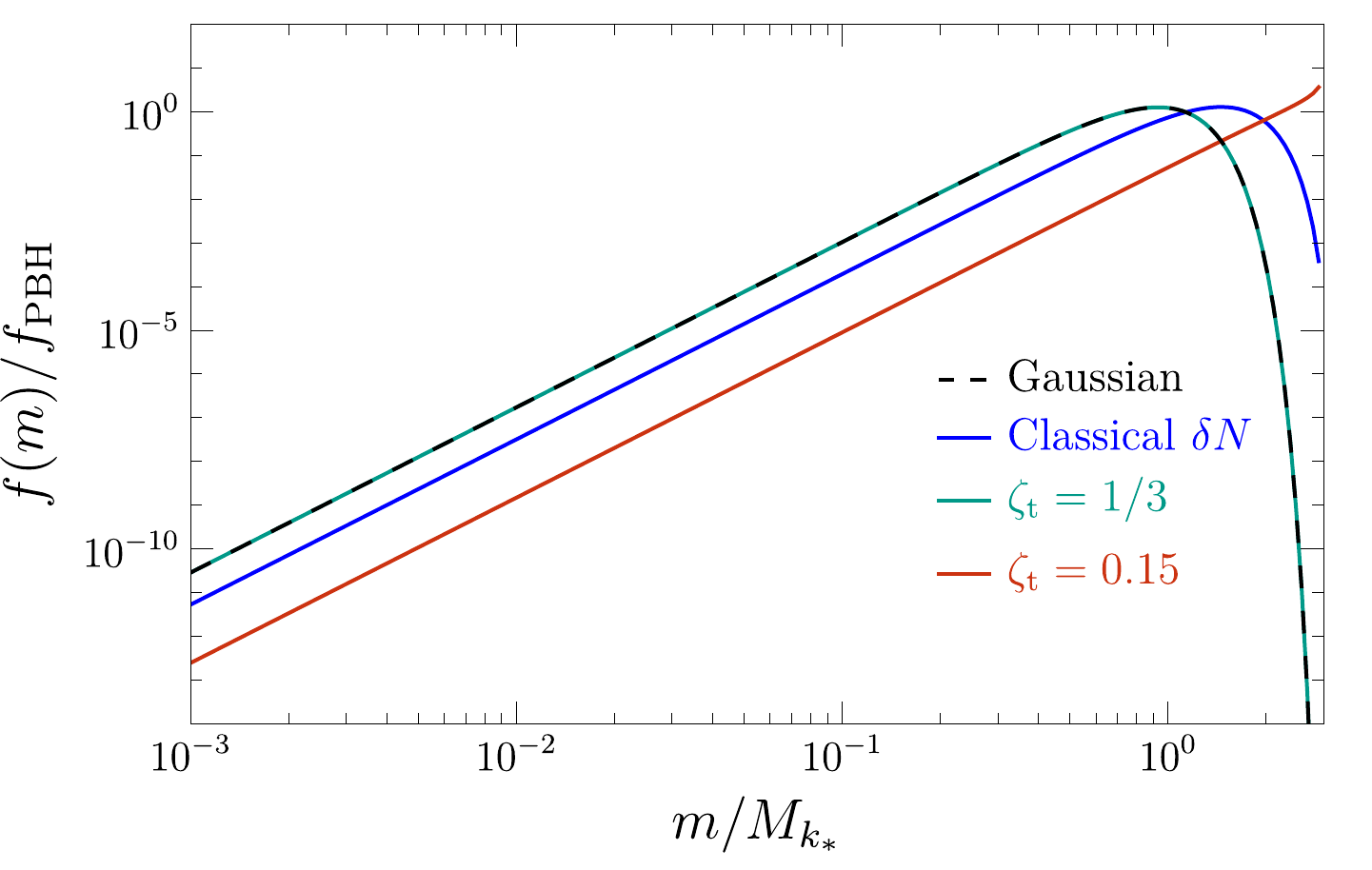}\includegraphics[width=0.5\textwidth]{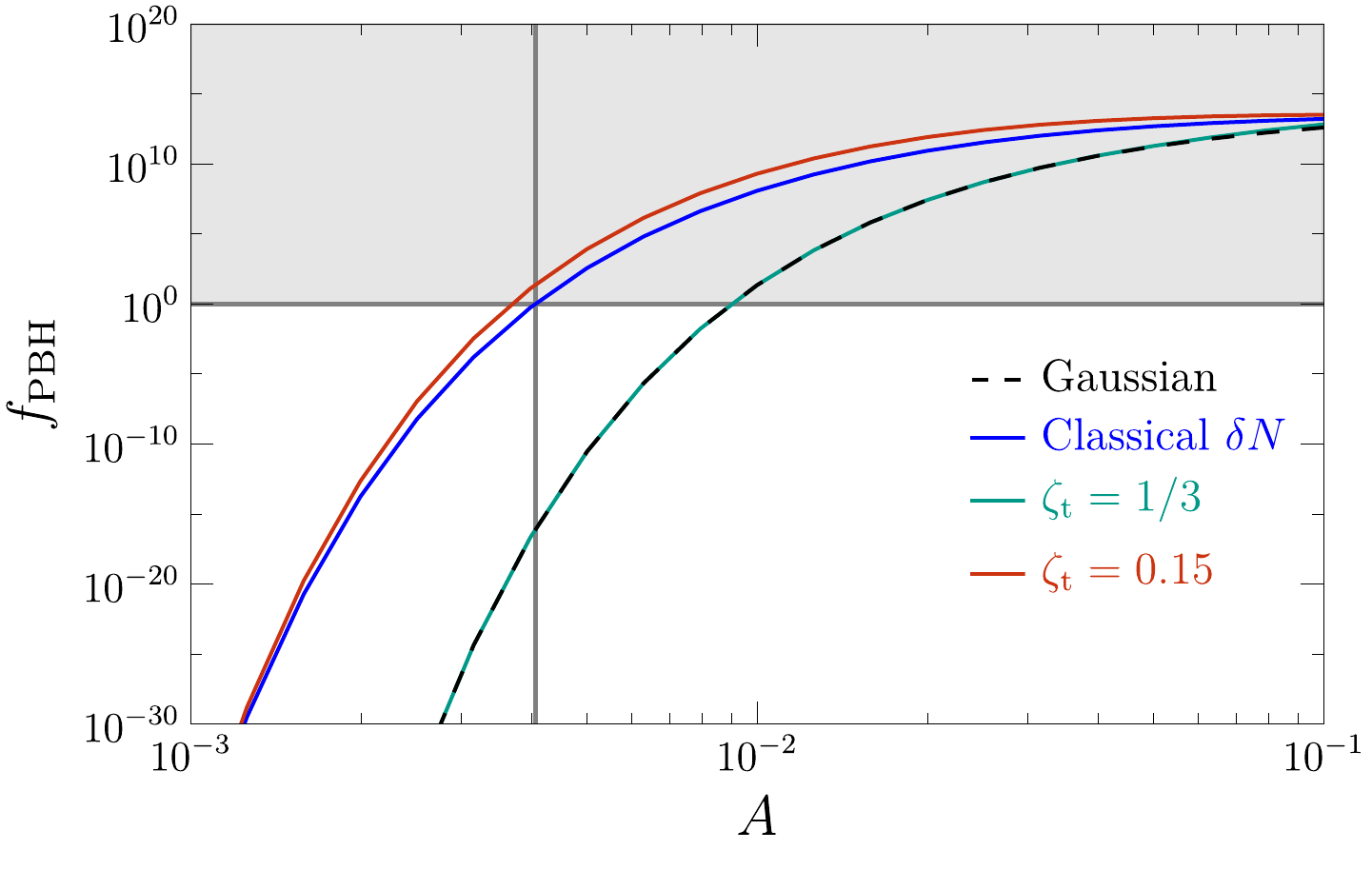}
\caption{Comparison of the piecewise exponential and the \cdN\ transformation. The slope, $\alpha=3$, is taken to match the \cdN\ tail. The two values of $\zeta_{\mathrm{t}}$ are chosen to approximately match the \cdN\ $P(\zeta)$ tail ($\zeta_{\mathrm{t}} = 1/3$) and $f_\PBH$ curve ($\zeta_{\mathrm{t}} = 0.15$). In the first three panels, $A \sim 4\times10^{-3}$, for which the \cdN\ distribution gives $f_\PBH = 1$. The blue shaded region in the top right panel represents the region relevant for PBH formation. The grey shaded region in the bottom right panel indicates overproduction of PBHs ($f_\PBH > 1$).}
\label{fig:Transition-vs-tail}
\end{figure}

Figure~\ref{fig:Transition-vs-tail} shows a comparison of the piecewise exponential model in \Eq{eq:P(zeta)-piecewise} with the Gaussian and \cdN\ cases. We choose $\alpha = 3$ in order to match the slope in the exponential tail of the \cdN\ distribution \eqref{eq:USRPzeta}. We select two values for the piecewise transition: $\zeta_{\mathrm{t}}=1/3$, to match the far tail of the \cdN\ distribution as can be seen in the top-left panel, and $\zeta_{\mathrm{t}}=0.15$, to approximately match the $f_\PBH$ curve in the bottom-right panel. The top-right panel shows $P(C_\uell)$, calculated using \Eq{eq:P(C_l)-main}. It is clear that for $\zeta_{\mathrm{t}} = 1/3$, there is no enhancement in the piecewise model over the Gaussian case, indicating that the exponential tail in the \cdN\ model does not actually enhance the production of PBHs, with the enhancement instead coming from the transition region between $\zeta \approx 0.2$ and $1.6$ in the top left panel. This is corroborated by the $f_\PBH$ curves, where the $\zeta_{\mathrm{t}} = 1/3$ case is identical to the Gaussian. It is also clear from the top right panel that the \cdN\ model, and the piecewise model with $\zeta_{\mathrm{t}} = 0.15$ chosen to produce a similar PBH enhancement, have significant support for $C_\uell$ values larger than the cutoff of 4/3. This suggests that if type II perturbations do collapse to form PBHs, they will contribute significantly to the total abundance in these non-Gaussian models.

The bottom left panel shows the mass distribution $f(m)$, normalised by the total dark matter fraction $f_\PBH$. As expected, the Gaussian and $\zeta_{\mathrm{t}} = 1/3$ distributions are identical. However, it is clear that the $\zeta_{\mathrm{t}} = 0.15$ curve is significantly different, without the decaying behaviour for $m>M_{k_*}$. This is because the mass distribution defined in \Eq{eq:f(m)} has a divergence at the mass corresponding to $C_\uell = 4/3$, coming from the $(1 - \frac{3}{4}C_\uell)$ term in the denominator. All of the mass distributions in fig.~\ref{fig:Transition-vs-tail} have this divergence at the right hand limit of the mass range, but it is not visible for the other curves, because $P(C_\uell)$ is decaying faster than the term in the denominator. However, for the $\zeta_{\mathrm{t}} = 0.15$ case, the top-right panel shows that $P(C_\uell)$ is decaying very slowly, so does not beat the diverging term. The mass distribution $f(m)$ arising from the \cdN\ model does decay for $m>M_{k_*}$, despite having a far tail slope of $\alpha = 3$ in $P(\zeta)$ like the piecewise model. This is because, as stated previously, the relevant part of $P(\zeta)$ is not the far tail, but the transition, during which the effective value of $\alpha$ is significantly larger, leading to a decaying $P(C_\uell)$ and a tail in the mass distribution. It should also be noted that the critical collapse relation in \Eq{eq:Critical-collapse} has been used right up to $C_\uell = 4/3$, but is expected to break down before this point \cite{Escriva:2020_Simulation}.

\begin{figure}[H]
\centering
\includegraphics[width=0.5\textwidth]{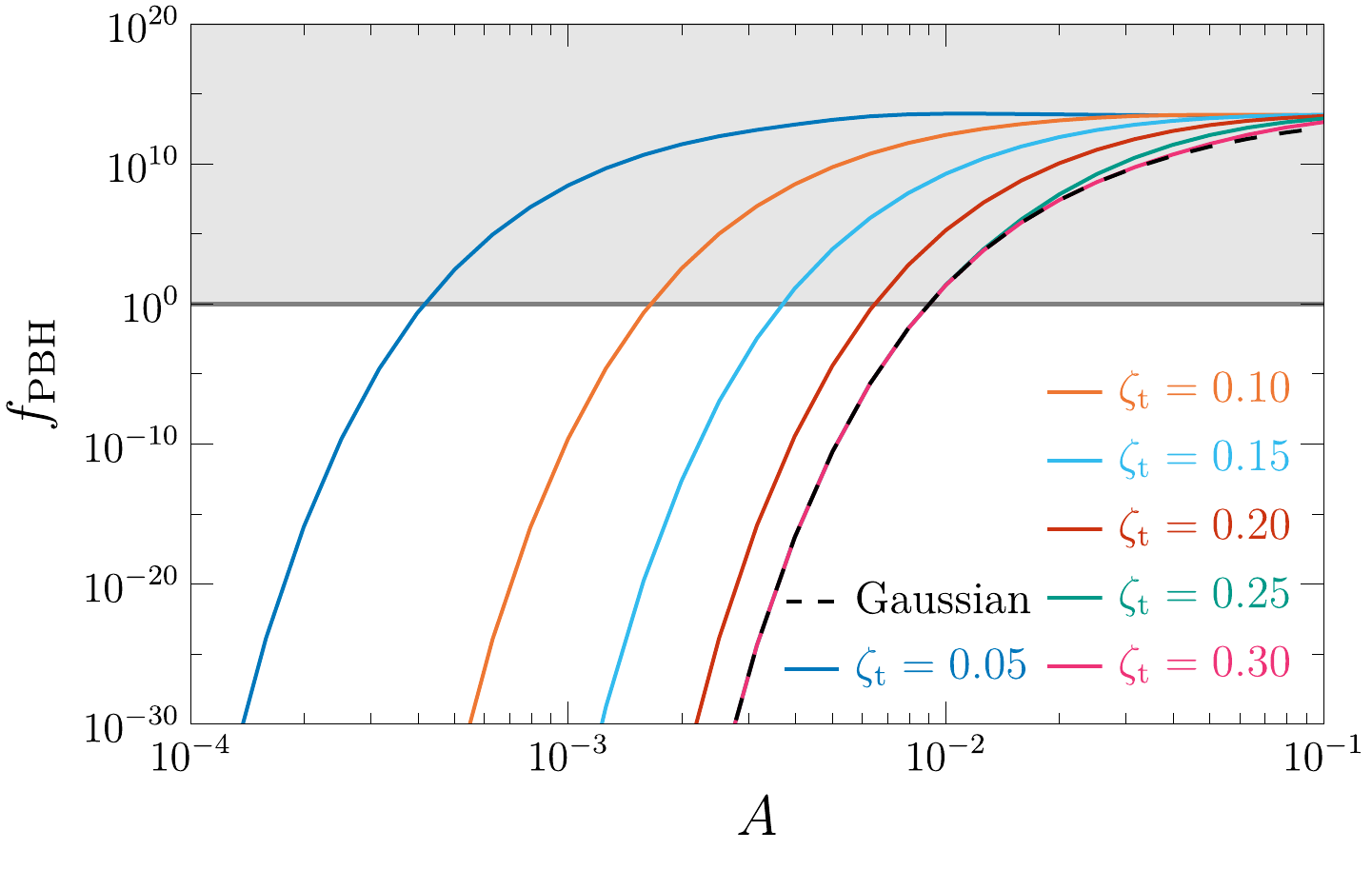}\includegraphics[width=0.5\textwidth]{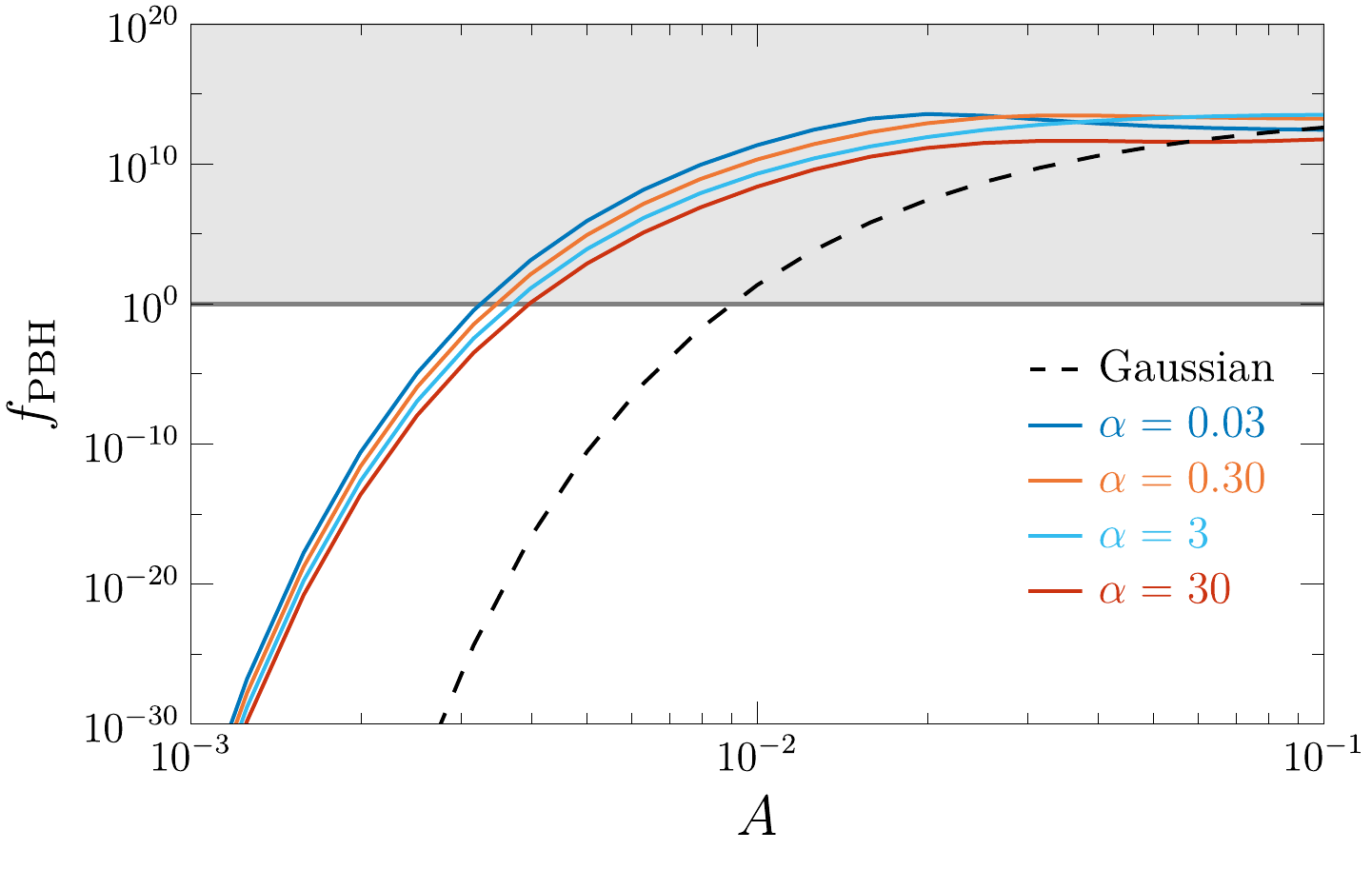}
\includegraphics[width=0.5\textwidth]{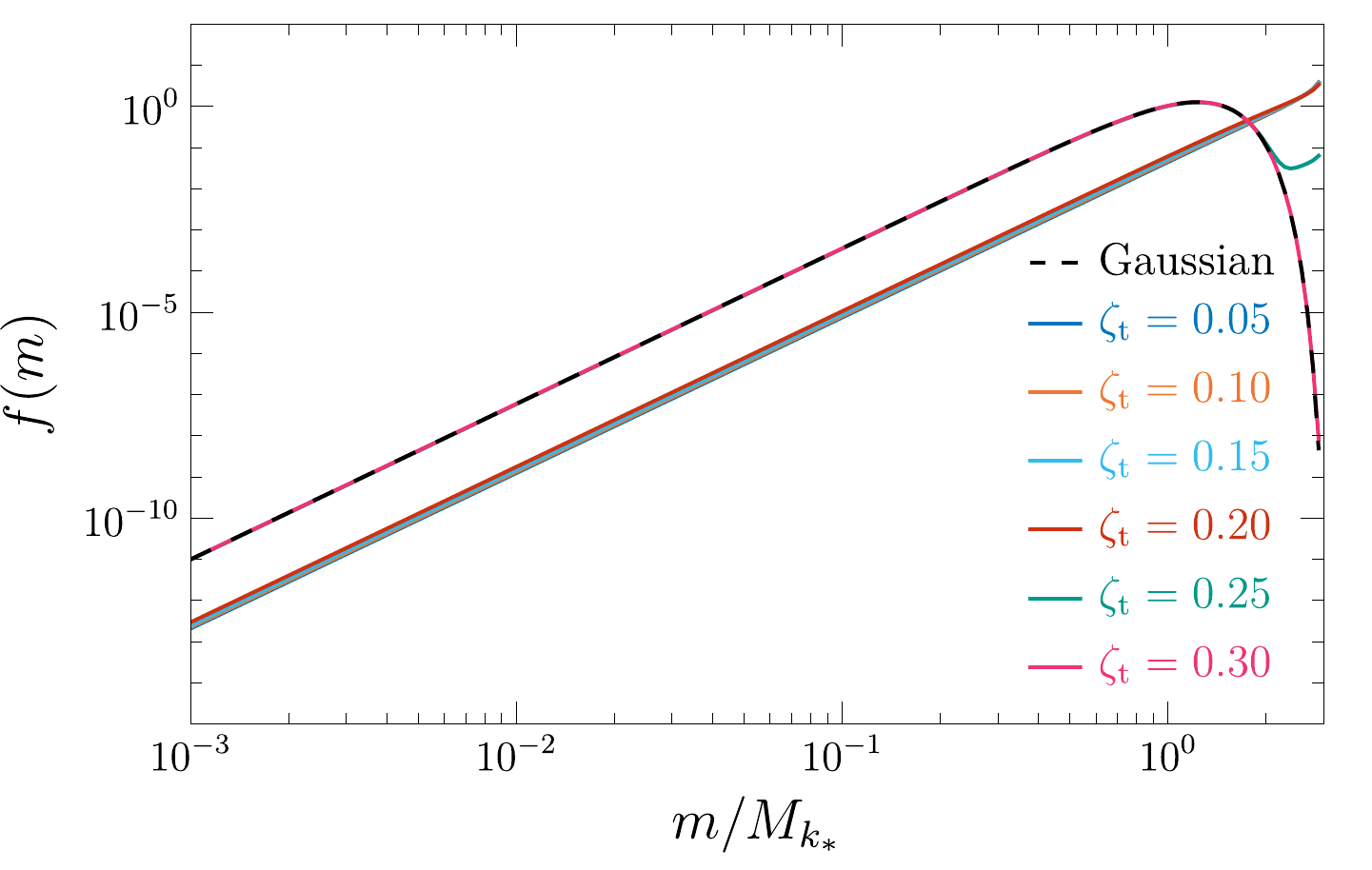}\includegraphics[width=0.5\textwidth]{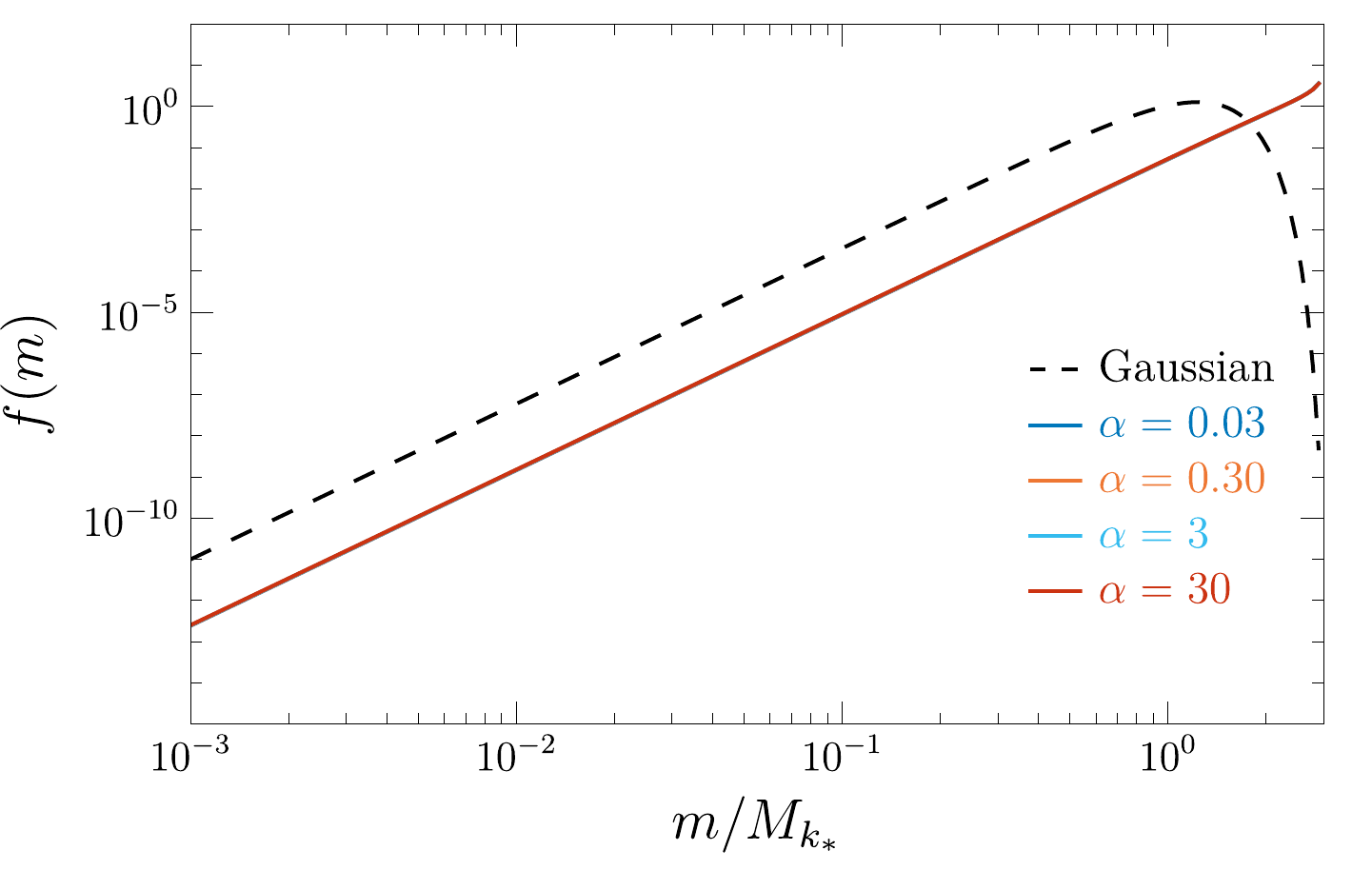}
\caption{Comparison of $f_\PBH(A)$ (top) and $f(m)$ (bottom) for different values of the piecewise distribution parameters $\zeta_{\mathrm{t}}$ and $\alpha$, with $\alpha = 3$ (left) and $\zeta_{\mathrm{t}} = 0.15$ (right). The grey shaded region in the top row indicates overproduction of PBHs ($f_\PBH > 1$). The mass distributions $f(m)$ in the bottom row are normalised to $f_\PBH = 1$.}
\label{fig:Piecewise-fPBH(A)-f(M)}
\end{figure}

Figure~\ref{fig:Piecewise-fPBH(A)-f(M)} shows the effect of changing the piecewise transition point $\zeta_{\mathrm{t}}$ and exponential slope $\alpha$ in \Eq{eq:P(zeta)-piecewise} on the total PBH abundance $f_\PBH$ and the mass distribution $f(m)$. As in fig.~\ref{fig:Transition-vs-tail}, the largest values of $\zeta_{\mathrm{t}}$ provide no enhancement over the Gaussian PDF except at the highest amplitudes, for which $f_\PBH \gg 1$. For smaller values of $\zeta_{\mathrm{t}}$, the exponential tail does provide a significant enhancement over the Gaussian case, greatly reducing the amplitude required for $f_\PBH = 1$. In contrast, changing $\alpha$ only moves the $f_\PBH$ curve by a (relatively) small amount, \eg by factors of $10$ rather than the $10^{10}$ coming from changing $\zeta_\text{t}$. The steeper values of $\alpha$ are suppressed compared to the shallower slopes, as expected. For large amplitudes, $f_\PBH$ is suppressed (even compared to the Gaussian). This suppression comes from the normalisation of $P(\zeta)$ when including a large tail. The details of this suppression are unimportant, since they correspond to $f_\PBH \gg 1$ and so these power spectrum amplitudes are disallowed anyway for PBH masses that survive to the present day.

In the bottom two panels we can see the effect of changing $\zeta_{\mathrm{t}}$ and $\alpha$ on the mass distribution $f(m)$. For $\zeta_{\mathrm{t}}$, there are three cases that are visible. The $\zeta_{\mathrm{t}} = 0.3$ curve lies on top of the Gaussian, as for the $\zeta_{\mathrm{t}} = 1/3$ case in fig.~\ref{fig:Transition-vs-tail}. For $\zeta_{\mathrm{t}} = 0.25$, the function $f(m)$ starts to decrease for $m>M_{k_*}$ as $P(C_\uell)$ decreases, before growing again due to the diverging term discussed previously. Finally, all values of $\zeta_{\mathrm{t}}$ smaller than 0.25 show no drop-off for $m>M_{k_*}$, instead being dominated by the diverging factor. Comparing the $f_\PBH$ and $f(m)$ plots for varying $\zeta_{\mathrm{t}}$, it seems clear that the amount of non-Gaussianity required to significantly reduce the power spectrum amplitude corresponding to a fixed $f_\PBH$ will also have a large impact on the mass distribution. This is important in the case of evading tight constraints such as those from $\mu$-distortions~\cite{Chluba:2012_Probing}. The right panel shows that, as with $f_\PBH$, the mass distribution is less affected by $\alpha$ than $\zeta_{\mathrm{t}}$, with all the curves appearing basically identical. This is consistent with the conclusion from fig.~\ref{fig:Transition-vs-tail} that the transition regime provides the enhancement, rather than the far tail.

\begin{table}[H]
\centering
\caption{Approximate values of $\nu_t = \zeta_{\mathrm{t}}/\sqrt{\Sigma_{YY}}$ and $P(\zeta_{\mathrm{t}})$ required for a specific $f_\PBH$, with a given $\alpha$. There is a linear relationship between $P(\zeta_{\mathrm{t}})$ and $f_\PBH$ for constant $\alpha$, and an inverse linear relationship between $f_\PBH$ and $\alpha$ for fixed $\nu_t$, which comes about from the integration of the far tail exponential behaviour.}
\label{tab:nut-values}
\begin{tabular}{c|c|c|c}
\textbf{$\alpha$} & \textbf{$\nu_t$} & \textbf{$P(\zeta_{\mathrm{t}})$} & \textbf{$f_\PBH$} \\ \hline
\multirow{2}{*}{3} & 8.2 & $10^{-10}$ & 1 \\
& 10.7 & $10^{-20}$ & $10^{-10}$ \\
0.3 & 8.2 & $10^{-10}$ & 10 \\
30 & 8.2 & $10^{-10}$ & $10^{-1}$
\end{tabular}
\end{table}

\section{Conclusions}

In this work, we have developed a non-perturbative treatment of primordial non-Gaussianity, when it is of a generalised local type,
and applied it to the formation of primordial black holes by relating the curvature perturbation $\zeta$ to the compaction function $C$. A non-perturbative method is essential when considering non-Gaussianities that are not described by the usual $\fnl$ expansion, such as those appearing in the stochastic treatment of inflation. Our method assumes there exists a generalised local transformation from a Gaussian field, $\zeta(\zeta_{\G}, r)$, and it remains to be seen whether this is a good description of the non-Gaussianity generated by stochastic diffusion.
Previous works have used the \cdN\ relation for a local-type non-Gaussian $\zeta(\zeta_{\G})$ with a PDF that smoothly transitions from a Gaussian to an exponential tail. By comparing to a simple piecewise model with a sudden transition between a Gaussian and an exponential with the same slope, we have shown that the enhanced PBH formation in the \cdN\ model is due to the transition regime, rather than the far tail. Additionally, we have examined the transition value, $\zeta_{\mathrm{t}}$, required for $f_\PBH = 1$ using our piecewise model. We find that $\nu_t = 
\zeta_{\mathrm{t}}/\sqrt{\Sigma_{YY}}$ is approximately independent of the peak amplitude of the Gaussian power spectrum, $A$, for fixed $f_\PBH$ and $\alpha$. We also see a linear relationship between $P(\zeta_{\mathrm{t}})$ and $f_\PBH$, and when changing the exponential slope $\alpha$ we find an inverse linear relation between $f_\PBH$ and $\alpha$, see table~\ref{tab:nut-values}. We also show that, for sufficiently shallow exponential tail slopes, the PBH mass distribution is dominated by a diverging term for high masses. This effect will be important when significant non-Gaussianity is required to reduce the power spectrum amplitude by orders of magnitude at fixed $f_\PBH$, \eg to evade the $\mu$-distortion constraints for supermassive PBHs.

\section*{Note added}
During the writing up of this paper, Ref.~\cite{Ferrante:2022_nG} appeared on the arXiv. The authors utilise the same mechanism of writing $\zeta$ in terms of $\zeta_{\G}$ and applying the product rule to the derivative in the compaction function, but restrict themselves to local transformations $\zeta(\zeta_{\G})$, rather than allowing $P(\zeta)$ to be arbitrary.

\section*{Acknowledgements}
The authors would like to thank Ilia Musco and Shi Pi for helpful discussions. This work is supported by the Science and Technology Facilities Council [grant numbers ST/S000550/1, ST/W001225/1, and ST/T506345/1]. For the purpose of open access, the authors have applied a Creative Commons Attribution (CC BY) licence to any Author Accepted Manuscript version arising. Supporting research data are available on reasonable request from the corresponding author, Andrew Gow.

\bibliographystyle{JHEP-edit} 
\bibliography{Non-perturbative_non-Gaussianity_paper}{}

\providecommand{\href}[2]{#2}\begingroup\raggedright\begin{thebibliography}{10}

\bibitem{Zel'dovich:1967_Cores}
{\relax Ya}.~B. Zel'dovich and I.~D. Novikov, \emph{{The Hypothesis of Cores
  Retarded during Expansion and the Hot Cosmological Model}},
  \href{https://ui.adsabs.harvard.edu/abs/1967SvA....10..602Z/abstract}{\emph{Soviet
  Astronomy} {\bfseries 10}{\bfseries (4)} (1967) 602}.

\bibitem{Hawking:1971_Gravitationally}
S.~Hawking, \emph{{Gravitationally Collapsed Objects of Very Low Mass}},
  \href{https://doi.org/10.1093/mnras/152.1.75}{\emph{Monthly Notices of the
  Royal Astronomical Society} {\bfseries 152}{\bfseries (1)} (1971) 75}.

\bibitem{Carr:1974_Black}
B.~J. Carr and S.~W. Hawking, \emph{{Black Holes in the Early Universe}},
  \href{https://doi.org/10.1093/mnras/168.2.399}{\emph{Monthly Notices of the
  Royal Astronomical Society} {\bfseries 168}{\bfseries (2)} (1974) 399}.

\bibitem{Carr:2020_Constraints}
B.~Carr, K.~Kohri, Y.~Sendouda and J.~Yokoyama, \emph{Constraints on primordial
  black holes}, \href{https://doi.org/10.1088/1361-6633/ac1e31}{\emph{Reports
  on Progress in Physics} {\bfseries 84}{\bfseries (11)} (2021) 116902}
  [\href{https://arxiv.org/abs/2002.12778}{{\ttfamily 2002.12778}}].

\bibitem{Bird:2016_GW}
S.~Bird \textit{et~al.}, \emph{{Did LIGO Detect Dark Matter?}},
  \href{https://doi.org/10.1103/PhysRevLett.116.201301}{\emph{Physical Review
  Letters} {\bfseries 116}{\bfseries (20)} (2016) 201301}
  [\href{https://arxiv.org/abs/1603.00464}{{\ttfamily 1603.00464}}].

\bibitem{Sasaki:GW}
M.~Sasaki, T.~Suyama, T.~Tanaka and S.~Yokoyama, \emph{{Primordial Black Hole
  Scenario for the Gravitational-Wave Event GW150914}},
  \href{https://doi.org/10.1103/PhysRevLett.117.061101}{\emph{Physical Review
  Letters} {\bfseries 117}{\bfseries (6)} (2016) 061101}
  [\href{https://arxiv.org/abs/1603.08338}{{\ttfamily 1603.08338}}].

\bibitem{Clesse:2017_Clustering}
S.~Clesse and J.~García-Bellido, \emph{{The clustering of massive Primordial
  Black Holes as Dark Matter: Measuring their mass distribution with advanced
  LIGO}}, \href{https://doi.org/10.1016/j.dark.2016.10.002}{\emph{Physics of
  the Dark Universe} {\bfseries 15} (2017) 142}
  [\href{https://arxiv.org/abs/1603.05234}{{\ttfamily 1603.05234}}].

\bibitem{Gow:2020_ACAJ}
A.~D. Gow, C.~T. Byrnes, A.~Hall and J.~A. Peacock, \emph{{Primordial black
  hole merger rates: distributions for multiple LIGO observables}},
  \href{https://doi.org/10.1088/1475-7516/2020/01/031}{\emph{Journal of
  Cosmology and Astroparticle Physics} {\bfseries 2020}{\bfseries (01)} (2020)
  031} [\href{https://arxiv.org/abs/1911.12685}{{\ttfamily 1911.12685}}].

\bibitem{Franciolini:2021_GW}
G.~Franciolini \textit{et~al.}, \emph{{Searching for a subpopulation of
  primordial black holes in LIGO-Virgo gravitational-wave data}},
  \href{https://doi.org/10.1103/PhysRevD.105.083526}{\emph{Physical Review D}
  {\bfseries 105}{\bfseries (8)} (2022) 083526}
  [\href{https://arxiv.org/abs/2105.03349}{{\ttfamily 2105.03349}}].

\bibitem{Planck:2018_Parameters}
{\scshape Planck} collaboration, \emph{{Planck 2018 results. VI. Cosmological
  parameters}},
  \href{https://doi.org/10.1051/0004-6361/201833910}{\emph{Astronomy \&
  Astrophysics} {\bfseries 641} (2020) A6}
  [\href{https://arxiv.org/abs/1807.06209}{{\ttfamily 1807.06209}}].

\bibitem{Gow:2021_ACPS}
A.~D. Gow, C.~T. Byrnes, P.~S. Cole and S.~Young, \emph{{The power spectrum on
  small scales: robust constraints and comparing PBH methodologies}},
  \href{https://doi.org/10.1088/1475-7516/2021/02/002}{\emph{Journal of
  Cosmology and Astroparticle Physics} {\bfseries 2021}{\bfseries (02)} (2021)
  002} [\href{https://arxiv.org/abs/2008.03289}{{\ttfamily 2008.03289}}].

\bibitem{Bullock:1996_Non-Gaussian}
J.~S. Bullock and J.~R. Primack, \emph{{Non-Gaussian fluctuations and
  primordial black holes from inflation}},
  \href{https://doi.org/10.1103/PhysRevD.55.7423}{\emph{Physical Review D}
  {\bfseries 55}{\bfseries (12)} (1997) 7423}
  [\href{https://arxiv.org/abs/astro-ph/9611106}{{\ttfamily
  astro-ph/9611106}}].

\bibitem{Young:2015_Influence}
S.~Young, D.~Regan and C.~T. Byrnes, \emph{Influence of large local and
  non-local bispectra on primordial black hole abundance},
  \href{https://doi.org/10.1088/1475-7516/2016/02/029}{\emph{Journal of
  Cosmology and Astroparticle Physics} {\bfseries 2016}{\bfseries (02)} (2016)
  029} [\href{https://arxiv.org/abs/1512.07224}{{\ttfamily 1512.07224}}].

\bibitem{Yoo:2019_Abundance}
C.-M. Yoo, J.-O. Gong and S.~Yokoyama, \emph{{Abundance of primordial black
  holes with local non-Gaussianity in peak theory}},
  \href{https://doi.org/10.1088/1475-7516/2019/09/033}{\emph{Journal of
  Cosmology and Astroparticle Physics} {\bfseries 2019}{\bfseries (09)} (2019)
  033} [\href{https://arxiv.org/abs/1906.06790}{{\ttfamily 1906.06790}}].

\bibitem{Palma:2020_Non-Gaussian}
G.~A. Palma, B.~Scheihing~Hitschfeld and S.~Sypsas, \emph{{Non-Gaussian CMB and
  LSS statistics beyond polyspectra}},
  \href{https://doi.org/10.1088/1475-7516/2020/02/027}{\emph{Journal of
  Cosmology and Astroparticle Physics} {\bfseries 2020}{\bfseries (02)} (2020)
  027} [\href{https://arxiv.org/abs/1907.05332}{{\ttfamily 1907.05332}}].

\bibitem{Taoso:2021_Non-gaussianities}
M.~Taoso and A.~Urbano, \emph{Non-gaussianities for primordial black hole
  formation},
  \href{https://doi.org/10.1088/1475-7516/2021/08/016}{\emph{Journal of
  Cosmology and Astroparticle Physics} {\bfseries 2021}{\bfseries (08)} (2021)
  016} [\href{https://arxiv.org/abs/2102.03610}{{\ttfamily 2102.03610}}].

\bibitem{Young:2022_non-G}
S.~Young, \emph{Peaks and primordial black holes: the effect of
  non-gaussianity},
  \href{https://doi.org/10.1088/1475-7516/2022/05/037}{\emph{Journal of
  Cosmology and Astroparticle Physics} {\bfseries 2022}{\bfseries (05)} (2022)
  037} [\href{https://arxiv.org/abs/2201.13345}{{\ttfamily 2201.13345}}].

\bibitem{Lyth:2005_Inflationary}
D.~H. Lyth and Y.~Rodríguez, \emph{{Inflationary Prediction for Primordial
  Non-Gaussianity}},
  \href{https://doi.org/10.1103/PhysRevLett.95.121302}{\emph{Physical Review
  Letters} {\bfseries 95}{\bfseries (12)} (2005) 121302}
  [\href{https://arxiv.org/abs/astro-ph/0504045}{{\ttfamily
  astro-ph/0504045}}].

\bibitem{Fujita:2013_Algorithm}
T.~Fujita, M.~Kawasaki, Y.~Tada and T.~Takesako, \emph{A new algorithm for
  calculating the curvature perturbations in stochastic inflation},
  \href{https://doi.org/10.1088/1475-7516/2013/12/036}{\emph{Journal of
  Cosmology and Astroparticle Physics} {\bfseries 2013}{\bfseries (12)} (2013)
  036} [\href{https://arxiv.org/abs/1308.4754}{{\ttfamily 1308.4754}}].

\bibitem{Vennin:2015_Correlation}
V.~Vennin and A.~A. Starobinsky, \emph{Correlation functions in stochastic
  inflation}, \href{https://doi.org/10.1140/epjc/s10052-015-3643-y}{\emph{The
  European Physical Journal C} {\bfseries 75} (2015) 413}
  [\href{https://arxiv.org/abs/1506.04732}{{\ttfamily 1506.04732}}].

\bibitem{Pattison:2017_Quantum}
C.~Pattison, V.~Vennin, H.~Assadullahi and D.~Wands, \emph{Quantum diffusion
  during inflation and primordial black holes},
  \href{https://doi.org/10.1088/1475-7516/2017/10/046}{\emph{Journal of
  Cosmology and Astroparticle Physics} {\bfseries 2017}{\bfseries (10)} (2017)
  046} [\href{https://arxiv.org/abs/1707.00537}{{\ttfamily 1707.00537}}].

\bibitem{Ezquiaga:2019_Exponential}
J.~M. Ezquiaga, J.~García-Bellido and V.~Vennin, \emph{The exponential tail of
  inflationary fluctuations: consequences for primordial black holes},
  \href{https://doi.org/10.1088/1475-7516/2020/03/029}{\emph{Journal of
  Cosmology and Astroparticle Physics} {\bfseries 2020}{\bfseries (03)} (2020)
  029} [\href{https://arxiv.org/abs/1912.05399}{{\ttfamily 1912.05399}}].

\bibitem{Vennin:2020_Thesis}
V.~Vennin, \emph{Stochastic inflation and primordial black holes}, Habilitation
  thesis, (2020) [\href{https://arxiv.org/abs/2009.08715}{{\ttfamily
  2009.08715}}].

\bibitem{Figueroa:2020_Non-Gaussian}
D.~G. Figueroa, S.~Raatikainen, S.~Räsänen and E.~Tomberg,
  \emph{{Non-Gaussian Tail of the Curvature Perturbation in Stochastic
  Ultraslow-Roll Inflation: Implications for Primordial Black Hole
  Production}},
  \href{https://doi.org/10.1103/PhysRevLett.127.101302}{\emph{Physical Review
  Letters} {\bfseries 127}{\bfseries (10)} (2021) 101302}
  [\href{https://arxiv.org/abs/2012.06551}{{\ttfamily 2012.06551}}].

\bibitem{Ando:2020_Power}
K.~Ando and V.~Vennin, \emph{Power spectrum in stochastic inflation},
  \href{https://doi.org/10.1088/1475-7516/2021/04/057}{\emph{Journal of
  Cosmology and Astroparticle Physics} {\bfseries 2021}{\bfseries (04)} (2021)
  057} [\href{https://arxiv.org/abs/2012.02031}{{\ttfamily 2012.02031}}].

\bibitem{Pattison:2021_USR}
C.~Pattison, V.~Vennin, D.~Wands and H.~Assadullahi, \emph{Ultra-slow-roll
  inflation with quantum diffusion},
  \href{https://doi.org/10.1088/1475-7516/2021/04/080}{\emph{Journal of
  Cosmology and Astroparticle Physics} {\bfseries 2021}{\bfseries (04)} (2021)
  080} [\href{https://arxiv.org/abs/2101.05741}{{\ttfamily 2101.05741}}].

\bibitem{Rigopoulos:2021_Inflation}
G.~Rigopoulos and A.~Wilkins, \emph{{Inflation is always semi-classical:
  diffusion domination overproduces Primordial Black Holes}},
  \href{https://doi.org/10.1088/1475-7516/2021/12/027}{\emph{Journal of
  Cosmology and Astroparticle Physics} {\bfseries 2021}{\bfseries (12)} (2021)
  027} [\href{https://arxiv.org/abs/2107.05317}{{\ttfamily 2107.05317}}].

\bibitem{Tada:2021_Statistics}
Y.~Tada and V.~Vennin, \emph{Statistics of coarse-grained cosmological fields
  in stochastic inflation},
  \href{https://doi.org/10.1088/1475-7516/2022/02/021}{\emph{Journal of
  Cosmology and Astroparticle Physics} {\bfseries 2022}{\bfseries (02)} (2022)
  021} [\href{https://arxiv.org/abs/2111.15280}{{\ttfamily 2111.15280}}].

\bibitem{Jackson:2022_Numerical}
J.~H.~P. Jackson \textit{et~al.}, \emph{Numerical simulations of stochastic
  inflation using importance sampling},
  \href{https://doi.org/10.1088/1475-7516/2022/10/067}{\emph{Journal of
  Cosmology and Astroparticle Physics} {\bfseries 2022}{\bfseries (10)} (2022)
  067} [\href{https://arxiv.org/abs/2206.11234}{{\ttfamily 2206.11234}}].

\bibitem{Animali:2022_Primordial}
C.~Animali and V.~Vennin, \emph{Primordial black holes from stochastic
  tunnelling},
  \href{https://doi.org/10.1088/1475-7516/2023/02/043}{\emph{Journal of
  Cosmology and Astroparticle Physics} {\bfseries 2023}{\bfseries (02)} (2023)
  043} [\href{https://arxiv.org/abs/2210.03812}{{\ttfamily 2210.03812}}].

\bibitem{Panagopoulos:2019_Primordial}
G.~Panagopoulos and E.~Silverstein, \emph{{Primordial Black Holes from
  non-Gaussian tails}},  (2019)
  [\href{https://arxiv.org/abs/1906.02827}{{\ttfamily 1906.02827}}].

\bibitem{Achucarro:2021_Hand-made}
A.~Achúcarro, S.~Céspedes, A.-C. Davis and G.~A. Palma, \emph{The hand-made
  tail: non-perturbative tails from multifield inflation},
  \href{https://doi.org/10.1007/JHEP05(2022)052}{\emph{Journal of High Energy
  Physics} {\bfseries 2022}{\bfseries (05)} (2022) 052}
  [\href{https://arxiv.org/abs/2112.14712}{{\ttfamily 2112.14712}}].

\bibitem{Cai:2022_Highly}
Y.-F. Cai \textit{et~al.}, \emph{{Highly non-Gaussian tails and primordial
  black holes from single-field inflation}},
  \href{https://doi.org/10.1088/1475-7516/2022/12/034}{\emph{Journal of
  Cosmology and Astroparticle Physics} {\bfseries 2022}{\bfseries (12)} (2022)
  034} [\href{https://arxiv.org/abs/2207.11910}{{\ttfamily 2207.11910}}].

\bibitem{Shibata:1999_Compaction}
M.~Shibata and M.~Sasaki, \emph{{Black hole formation in the Friedmann
  universe: Formulation and computation in numerical relativity}},
  \href{https://doi.org/10.1103/PhysRevD.60.084002}{\emph{Physical Review D}
  {\bfseries 60}{\bfseries (8)} (1999) 084002}
  [\href{https://arxiv.org/abs/gr-qc/9905064}{{\ttfamily gr-qc/9905064}}].

\bibitem{Harada:2015_Compaction}
T.~Harada, C.-M. Yoo, T.~Nakama and Y.~Koga, \emph{Cosmological long-wavelength
  solutions and primordial black hole formation},
  \href{https://doi.org/10.1103/PhysRevD.91.084057}{\emph{Physical Review D}
  {\bfseries 91}{\bfseries (8)} (2015) 084057}
  [\href{https://arxiv.org/abs/1503.03934}{{\ttfamily 1503.03934}}].

\bibitem{Kopp:2010_Separate}
M.~Kopp, S.~Hofmann and J.~Weller, \emph{Separate universes do not constrain
  primordial black hole formation},
  \href{https://doi.org/10.1103/PhysRevD.83.124025}{\emph{Physical Review D}
  {\bfseries 83}{\bfseries (12)} (2011) 124025}
  [\href{https://arxiv.org/abs/1012.4369}{{\ttfamily 1012.4369}}].

\bibitem{DeLuca:2022_Note}
V.~De~Luca and A.~Riotto, \emph{{A note on the abundance of primordial black
  holes: Use and misuse of the metric curvature perturbation}},
  \href{https://doi.org/10.1016/j.physletb.2022.137035}{\emph{Physics Letters
  B} {\bfseries 828} (2022) 137035}
  [\href{https://arxiv.org/abs/2201.09008}{{\ttfamily 2201.09008}}].

\bibitem{Biagetti:2021_Formation}
M.~Biagetti \textit{et~al.}, \emph{The formation probability of primordial
  black holes},
  \href{https://doi.org/10.1016/j.physletb.2021.136602}{\emph{Physics Letters
  B} {\bfseries 820} (2021) 136602}
  [\href{https://arxiv.org/abs/2105.07810}{{\ttfamily 2105.07810}}].

\bibitem{Kitajima:2021_Primordial}
N.~Kitajima, Y.~Tada, S.~Yokoyama and C.-M. Yoo, \emph{{Primordial black holes
  in peak theory with a non-Gaussian tail}},
  \href{https://doi.org/10.1088/1475-7516/2021/10/053}{\emph{Journal of
  Cosmology and Astroparticle Physics} {\bfseries 2021}{\bfseries (10)} (2021)
  053} [\href{https://arxiv.org/abs/2109.00791}{{\ttfamily 2109.00791}}].

\bibitem{Choptuik:1992_Critical-collapse}
M.~W. Choptuik, \emph{Universality and scaling in gravitational collapse of a
  massless scalar field},
  \href{https://doi.org/10.1103/PhysRevLett.70.9}{\emph{Physical Review
  Letters} {\bfseries 70}{\bfseries (1)} (1993) 9}.

\bibitem{Evans:1994_Critical-collapse}
C.~R. Evans and J.~S. Coleman, \emph{Critical phenomena and self-similarity in
  the gravitational collapse of radiation fluid},
  \href{https://doi.org/10.1103/PhysRevLett.72.1782}{\emph{Physical Review
  Letters} {\bfseries 72}{\bfseries (12)} (1994) 1782}
  [\href{https://arxiv.org/abs/gr-qc/9402041}{{\ttfamily gr-qc/9402041}}].

\bibitem{Niemeyer:1998_Critical-collapse}
J.~C. Niemeyer and K.~Jedamzik, \emph{{Near-Critical Gravitational Collapse and
  the Initial Mass Function of Primordial Black Holes}},
  \href{https://doi.org/10.1103/PhysRevLett.80.5481}{\emph{Physical Review
  Letters} {\bfseries 80}{\bfseries (25)} (1998) 5481}
  [\href{https://arxiv.org/abs/astro-ph/9709072}{{\ttfamily
  astro-ph/9709072}}].

\bibitem{Musco:2018_CC}
I.~Musco, \emph{{Threshold for primordial black holes: Dependence on the shape
  of the cosmological perturbations}},
  \href{https://doi.org/10.1103/PhysRevD.100.123524}{\emph{Physical Review D}
  {\bfseries 100}{\bfseries (12)} (2019) 123524}
  [\href{https://arxiv.org/abs/1809.02127}{{\ttfamily 1809.02127}}].

\bibitem{Young:2019_CC}
S.~Young, \emph{{The primordial black hole formation criterion re-examined:
  Parametrisation, timing and the choice of window function}},
  \href{https://doi.org/10.1142/S0218271820300025}{\emph{International Journal
  of Modern Physics D} {\bfseries 29}{\bfseries (02)} (2019) 2030002}
  [\href{https://arxiv.org/abs/1905.01230}{{\ttfamily 1905.01230}}].

\bibitem{Germani:2018_CC}
C.~Germani and I.~Musco, \emph{{Abundance of Primordial Black Holes Depends on
  the Shape of the Inflationary Power Spectrum}},
  \href{https://doi.org/10.1103/PhysRevLett.122.141302}{\emph{Physical Review
  Letters} {\bfseries 122}{\bfseries (14)} (2019) 141302}
  [\href{https://arxiv.org/abs/1805.04087}{{\ttfamily 1805.04087}}].

\bibitem{Germani:2019_Nonlinear}
C.~Germani and R.~K. Sheth, \emph{{Nonlinear statistics of primordial black
  holes from Gaussian curvature perturbations}},
  \href{https://doi.org/10.1103/PhysRevD.101.063520}{\emph{Physical Review D}
  {\bfseries 101}{\bfseries (6)} (2020) 063520}
  [\href{https://arxiv.org/abs/1912.07072}{{\ttfamily 1912.07072}}].

\bibitem{Escriva:2020_CC}
A.~Escrivà, C.~Germani and R.~K. Sheth, \emph{Analytical thresholds for black
  hole formation in general cosmological backgrounds},
  \href{https://doi.org/10.1088/1475-7516/2021/01/030}{\emph{Journal of
  Cosmology and Astroparticle Physics} {\bfseries 2021}{\bfseries (01)} (2021)
  030} [\href{https://arxiv.org/abs/2007.05564}{{\ttfamily 2007.05564}}].

\bibitem{Musco:2020_274}
I.~Musco, V.~De~Luca, G.~Franciolini and A.~Riotto, \emph{{Threshold for
  primordial black holes. II. A simple analytic prescription}},
  \href{https://doi.org/10.1103/PhysRevD.103.063538}{\emph{Physical Review D}
  {\bfseries 103}{\bfseries (6)} (2021) 063538}
  [\href{https://arxiv.org/abs/2011.03014}{{\ttfamily 2011.03014}}].

\bibitem{Cai:2018_Revisiting}
Y.-F. Cai \textit{et~al.}, \emph{{Revisiting non-Gaussianity from non-attractor
  inflation models}},
  \href{https://doi.org/10.1088/1475-7516/2018/05/012}{\emph{Journal of
  Cosmology and Astroparticle Physics} {\bfseries 2018}{\bfseries (05)} (2018)
  012} [\href{https://arxiv.org/abs/1712.09998}{{\ttfamily 1712.09998}}].

\bibitem{Atal:2019_Primordial}
V.~Atal, J.~Garriga and A.~Marcos-Caballero, \emph{{Primordial black hole
  formation with non-Gaussian curvature perturbations}},
  \href{https://doi.org/10.1088/1475-7516/2019/09/073}{\emph{Journal of
  Cosmology and Astroparticle Physics} {\bfseries 2019}{\bfseries (09)} (2019)
  073} [\href{https://arxiv.org/abs/1905.13202}{{\ttfamily 1905.13202}}].

\bibitem{Pi:2021_Primordial}
S.~Pi and M.~Sasaki, \emph{{Primordial Black Hole Formation in Non-Minimal
  Curvaton Scenario}},  (2021)
  [\href{https://arxiv.org/abs/2112.12680}{{\ttfamily 2112.12680}}].

\bibitem{Escriva:2020_Simulation}
A.~Escrivà, \emph{Simulation of primordial black hole formation using
  pseudo-spectral methods},
  \href{https://doi.org/10.1016/j.dark.2020.100466}{\emph{Physics of the Dark
  Universe} {\bfseries 27} (2020) 100466}
  [\href{https://arxiv.org/abs/1907.13065}{{\ttfamily 1907.13065}}].

\bibitem{Chluba:2012_Probing}
J.~Chluba, A.~L. Erickcek and I.~Ben-Dayan, \emph{{PROBING THE INFLATON:
  SMALL-SCALE POWER SPECTRUM CONSTRAINTS FROM MEASUREMENTS OF THE COSMIC
  MICROWAVE BACKGROUND ENERGY SPECTRUM}},
  \href{https://doi.org/10.1088/0004-637x/758/2/76}{\emph{The Astrophysical
  Journal} {\bfseries 758}{\bfseries (2)} (2012) 76}
  [\href{https://arxiv.org/abs/1203.2681}{{\ttfamily 1203.2681}}].

\bibitem{Ferrante:2022_nG}
G.~Ferrante, G.~Franciolini, A.~J. Iovino and A.~Urbano, \emph{{Primordial
  non-Gaussianity up to all orders: Theoretical aspects and implications for
  primordial black hole models}},
  \href{https://doi.org/10.1103/PhysRevD.107.043520}{\emph{Phys. Rev. D}
  {\bfseries 107}{\bfseries (4)} (2023) 043520}
  [\href{https://arxiv.org/abs/2211.01728}{{\ttfamily 2211.01728}}].

\end{thebibliography}\endgroup

\end{document}